\documentclass[twocolumn,superscriptaddress,nofootinbib,amsmath,amssymb,amsfonts]{aastex631}

\usepackage{graphicx}%
\usepackage{dcolumn}%
\usepackage{enumitem}%
\usepackage{bm}%
\usepackage{hyperref}%
\hypersetup{
  colorlinks=true,
  linkcolor=blue,
  citecolor=cyan,
  urlcolor=cyan
}
\usepackage{colortbl}
\usepackage{amsmath}
\usepackage{amssymb}
\usepackage{mathrsfs}
\usepackage{times}

\newcommand{\cf}{cf.~}
\newcommand{\ie}{i.e.,~}
\newcommand{\eg}{e.g.,~}

\begin{document}

\title{Self-consistent multidimensional Penrose process driven by magnetic reconnection}

\author[0000-0003-0412-0491]{Filippo Camilloni}
\affiliation{Institut f\"ur Theoretische Physik, Goethe Universit\"at, Max-von-Laue-Stra{\ss}e 1, 60438 Frankfurt am Main, Germany}
\author[0000-0002-1330-7103]{Luciano Rezzolla}
\affiliation{Institut f\"ur Theoretische Physik, Goethe Universit\"at, Max-von-Laue-Stra{\ss}e 1, 60438 Frankfurt am Main, Germany}
\affiliation{School of Mathematics, Trinity College, Dublin 2, Ireland}
\affiliation{Frankfurt Institute for Advanced Studies, Ruth-Moufang-Str. 1, 60438 Frankfurt am Main, Germany}

\date{\today}

\begin{abstract}
Astronomical observations and numerical simulations are providing
increasing evidence that resistive effects in plasmas around black holes
play an important role in determining the phenomenology observed from
these objects. In this spirit, we present a general approach to the study
of a Penrose process driven by plasmoids that are produced at
reconnection sites along current sheets. Our formalism is meant to
determine the physical conditions that make a plasmoid-driven Penrose
process energetically viable and can be applied to scenarios that are
matter- or magnetic-field-dominated, that is, in magnetohydrodynamical or
force-free descriptions. By exploring reconnection from an axisymmetric
but curved surface, our approach can be considered genuinely
multidimensional and allows us to explore conditions that are beyond the
ones explored so far and that have been restricted to the equatorial
plane. Furthermore, it provides a direct contact with numerical
simulations of accretion onto black holes, which exhibit an intense
reconnection activity outside the equatorial plane. Finally, to describe
the kinematics of the plasma self-consistently, we use the well-known
configuration of an equilibrium torus with a purely toroidal magnetic
field. For such a torus, we discuss the existence of an ``ergobelt'', \ie
a nontrivial surface penetrating the ergosphere and acting as a natural a
site for the occurrence of reconnection, and from where we estimate the
energetics of a plasmoid-driven Penrose process.
\end{abstract}


\section{Introduction}
%
In a resistive plasma, magnetic reconnection consists in a sudden
rearrangement of the magnetic-field topology caused by the local
interaction of field lines with opposite polarity. This process leads to
a rapid conversion of magnetic energy into thermal and kinetic energy of
two plasma outflows with locally isotropic properties and moving in
opposite directions at relativistic speeds, the plasmoids. The ubiquitous
and fundamental role played by magnetic reconnection in astrophysical
plasmas is now widely recognized on scales ranging from stellar flares
and coronal mass ejections~\citep{su2013imaging,
  annurev:/content/journals/10.1146/annurev-astro-082708-101726,
  RevModPhys.82.603}, to high-energy sources such as pulsar
magnetospheres~\citep{Uzdensky_2014}, accreting black holes
(BHs)~\citep{Beloborodov_2017, Ripperda2019, Akiyama2019_L5_etal,
  EHT_SgrA_PaperV_etal}, and binary neutron-star mergers leading to the
short gamma-ray burst phenomenology~\citep{Palenzuela:2008sf, Liu:2008xy,
  Anderson2008, Rezzolla:2011, Palenzuela2013a, Kiuchi2015,
  Dionysopoulou2015}.

Recent analytic studies, as well as numerical simulations, either in
magnetohydrodynamics (MHD) or with particle-in-cell (PIC) approaches,
have unfolded a rich phenomenology associated to magnetic reconnection
and plasmoid generation, thus enhancing our understanding of these
processes both in special-relativistic
regimes~\citep{Liu:2016sqd,PhysRevLett.113.045001,
  PhysRevLett.113.155005, PhysRevLett.114.115003,PhysRevLett.114.095002,
  Ball2018, Meringolo2023}, and when general-relativistic effects become
relevant~\citep{Koide:2008xr, Asenjo:2017gsv, Parfrey2019,
  PhysRevD.103.023014, Bransgrove2021, Fan:2024fcy, Shen2024}. In this
regard, astrophysical BHs constitute unique theoretical laboratories to
explore extreme conditions of plasma electrodynamics. Indeed, one the
most remarkable predictions of general relativity is the possibility of
extracting rotational energy of the BH by invoking the negative inflow of
energy and angular momentum at the horizon~\citep{Lasota:2013kia}, with
various processes differing only in the physical agent operating the
extraction. Notable examples range from single particles associated to
mechanical~\citep{Penrose:1971uk, Ruffini2024} and collisional Penrose
process (PP)~\citep{Banados2009, Bejger2012, PhysRevLett.113.261102,
  Berti2015b}, to super-radiant scalar fields~\citep{Press:1972zz,
  Pani2012, Bosch2016, East2017}, and to force-free electrodynamics (FFE)
fields in the Blandford--Znajek (BZ)
mechanism~\citep{Blandford:1977ds,Dadich2018}.
	
Given the presence of strong magnetic fields and relativistic plasmas
near astrophysical BHs, it is natural to expect a manifestation of the PP
triggered by magnetic reconnection and mediated by plasmoids. This idea
was originally proposed and explored analytically by Koide and
Arai~\citep{Koide:2008xr}, to be further reviewed and extended by Asenjo
and Comisso~\citep{Asenjo:2017gsv, PhysRevD.103.023014}, who computed the
power extracted and the efficiency of the process in terms of the local
reconnection rate~\citep[see also][for related PIC
  simulations]{Parfrey2019}. The basic picture of this process
involves
a current-sheet with large aspect-ratio forming within the BH
ergosphere. As ordinary in reconnection processes, such a current-sheet
fragments via the tearing instability into a chain of plasmoids (or
magnetic flux-tubes in three dimensions) that are accelerated away from
the reconnection layer, becoming an efficient channel to trasport
magnetic energy away from the reconnection site. In a steady state, pairs
of plasmoids are ejected from the reconnection site, where, for each
pair, a plasmoid will move outwards to large distances, while the other
one will be ingoing and move towards the event horizon. If the ingoing
plasmoid has a negative energy-at-infinity, it will drive a PP. It is
important to note that the plasmoid that is relevant for the PP is the
ingoing one and that it needs to travel only a lengthscale of
$\mathcal{O}(M)$ within the ergosphere to reach the event horizon and
extract energy via the PP. The details of what happens to the outgoing
plasmoid are not important for the success of the PP, and the outgoing
plasmoid could transmit its energy-at-infinity just outside the
ergosphere after being converted into electromagnetic bursts or at very
large distances (e.g., via synchrotron cooling\footnote{Radiative losses
are important in a realistic description of plasmoids propagating in the
surrounding plasma and we expect these losses to be particularly relevant
for those plasmoids travelling large distances of $\mathcal{O}(100 \,
M)$~\citep{Aimar2023} and for which synchrotron cooling can be
effective. Radiative losses will instead affect only mildly those
plasmoids that propagate from within the ergoregion to the event horizon,
as they travel much shorter lengths of $\mathcal{O}(M)$.}). The PP would
work in either case.

Despite the importance of the initial works by \citet{Koide:2008xr} and
\citet{PhysRevD.103.023014}, they contain a number of approximations that
prevent from testing them in numerical environments and assessing their
viability in realistic astrophysical scenarios. Arguably, the most
serious of these limitations rests with the lack of a self-consistent
description of the plasma dynamics, either in the MHD or FFE regimes. In
turn, this forces the use of number of assumptions on the dynamics of the
plasma that are either far from known solutions and numerical simulations
(see, \eg \citet{Mahlmann2020, Ripperda2020, Nathanail2020,
  Nathanail2020c, Ripperda2022, Crinquand2022, Dimitropoulos:2024iqz}),
or difficult to realise in practice.

To overcome these difficulties, it is important to construct reconnection
models that are consistent with well-known plasma solutions or that share
dynamical features and conditions that are similar to those encountered
in simulations. The goal of this work is therefore that of having
  an analytic understanding of the energetics of the PP mediated by
  plasmoid for a non-trivial distribution of matter and electromagnetic
  fields. We do this by exploiting a self-consistent, fully covariant
analytic description of plasma and fields of degenerate electrodynamics
that includes ideal MHD and FFE as particular examples. More
specifically, we analyse for the first time the conditions under which a
distribution of magnetised matter penetrating the ergosphere and not
restricted to the equatorial plane can produce plasmoids with negative
energy-at-infinity. In this way, we can study the extraction of energy
and angular momentum from a Kerr BH in a self-consistent scenario that is
not too far from the conditions explored in simulations. The
  ultimate expectation is that the insight that will be gained via these
  analytical calculations will be of help in understanding the results of
  the simulations of the plasmoid-driven PP once they will have reached a
  sufficient degree of realism.

\section{Degenerate electrodynamics around a Kerr BH}

We consider the electrodynamics of a Kerr BH with mass $M$ and specific
angular momentum $a$. Following~\citet{Thorne82}, and introducing
$\boldsymbol{F}$ and $\eta_{\mu\nu\rho\sigma}$ respectively as the
Faraday tensor and the Levi-Civita symbol, we refer to as
\emph{degenerate} those electromagnetic fields that satisfy the condition
$\eta_{\mu\nu\rho\sigma} F^{\mu\nu} F^{\rho\sigma}=0$ (or, equivalently,
satisfying the condition of orthogonality between the electric and
magnetic fields $\boldsymbol{E} \cdot \boldsymbol{B}=0$) and
\emph{magnetically dominated} if \ie $F_{\mu\nu} F^{\mu\nu}>0$ (or,
equivalently, if $B^2 - E^2 > 0$). Under this definition, force-free
electromagnetic fields are clearly degenerate, but degeneracy can occur
more generally and, indeed, an electromagnetic field can be degenerate
without being force-free. This is because degeneracy appears whenever the
electric field in the rest frame of the plasma vanishes, \ie when
$\boldsymbol{F} \cdot \boldsymbol{u}=0$, where $\boldsymbol{u}$ is the
plasma four-velocity. Clearly, a plasma with infinite conductivity, such
as that characterising the ideal-MHD limit, will have a zero comoving
electric field and hence will be degenerate but not force-free. As a
result, the conditions of a degenerate and magnetically dominated plasma
describe both the force-free and the ideal-MHD conditions~\citep[see
  also][for more recent discussions]{Gralla:2014yja, Chael:2023pwp,
  Mizuno2024}. The orthogonality of electric and magnetic fields and are
violated only on the microscopical scales of the current-sheets where
reconnection takes place~\citep{Liu:2016sqd}. Degeneracy also implies
that stationary and axisymmetric fields are expressed in terms of a
magnetic-flux function $\Psi=\Psi(r,\theta)$ that, under a suitable gauge
choice, coincides with the toroidal component of the electromagnetic
potential. As a result, constant-$\Psi$ surfaces determine the poloidal
magnetic fields, the poloidal current $I=I(r,\theta)$, which is
proportional to the toroidal field, and the angular velocity of the
magnetic-field lines $\Omega
=\Omega(\Psi)$~\citep{Gralla:2014yja}.\footnote{Despite retaining
axisymmetry as a working assumption, our approach is multidimensional, as
the current-sheets associated with such configurations are off-equatorial
and described by surfaces extending in the polar, azimuthal, and radial
direction. This is in contrast with the equatorial current-sheets that
only span the latter two.}
%

\subsection{Zero angular momentum observer frame}

A $3$+$1$ decomposition of spacetime has a long history in BH
electrodynamics~{\citep{Landau-Lifshitz2, Thorne82}}, since it provides
equations in a form that closely resembles the classical Maxwell
equations, and can then be employed in numerical simulations. In this
case, the metric can be written in terms the {lapse function} $\alpha$,
the {shift vector} $\beta^\phi$ and the {spatial metric}
$\gamma_{ij}$~\citep[see][and also the
  Appendix]{Rezzolla_book:2013}. Within this decomposition, a
particularly convenient class of observers with a long history of use in
the literature~\citep{Bardeen72} is that with Zero Angular Momentum
(ZAMOs) with worldline tangent
\begin{equation}
\boldsymbol{\eta}:= \eta^\mu\boldsymbol{\partial_\mu} =
\alpha^{-1}(\boldsymbol{\partial_t} + \omega_{_{\rm Z}}
\boldsymbol{\partial_\phi})\,,
\end{equation}
and whose angular velocity is $\omega_{_{\rm Z}} := -\beta^\phi$. Such an
observer carries a local orthonormal (Cartesian) tetrad,
$\boldsymbol{\hat e}_{(\alpha)}$ with $\alpha=T,X,Y,Z$, such that on the
equatorial plane ($\theta=\pi/2$), the $X$ and $Y$ legs align
respectively with the $r$ and $\phi$ directions, whereas the $Z$ leg is
orthogonal to the equatorial plane and parallel to the BH spin
(quantities in the ZAMO frame are marked with a ``hat'' and the
components with bracketed indices).

{Introducing $^*\boldsymbol{F}$ as the dual of the Faraday tensor, the}
magnetic field measured by the ZAMO,
$\hat{\mathcal{B}}^\mu:=-{^*F}^{\mu\nu}\eta_\nu$, is purely spacelike,
and in ZAMO-adapted coordinates one has that
$\boldsymbol{\hat{\mathcal{B}}} = \hat{\mathcal{B}}^{(\alpha)}
\boldsymbol{\hat e}_{(\alpha)}$ reads
\begin{equation}
    \boldsymbol{\hat{\mathcal{B}}}
    =\frac{\partial_\theta\Psi}{\sqrt{\Pi}\sin\theta} \boldsymbol{\hat
      e}_{(X)} - \frac{I}{\sqrt{\Delta}\sin\theta} \boldsymbol{\hat
      e}_{(Y)} + \frac{\sqrt{\Delta}\partial_r\Psi}{
      \sqrt{\Pi}\sin\theta} \boldsymbol{\hat e}_{(Z)}\,,
\end{equation}
where
\begin{align}
&\Delta := r^2-2M r +a^2\,, \\
&\Pi := (r^2+a^2)^2-a^2\Delta \sin^2\theta\,,
\end{align}
are two metric functions of the Kerr solution and with the positions of
the event horizons $r_{\pm}$ given by roots of $\Delta=(r-r_+)(r-r_-)$
(see also the Appendix).

The electric field, $\hat{\mathcal{E}}^\mu:=F^{\mu\nu}\eta_\nu$, is
orthogonal to $\boldsymbol{\hat{\mathcal{B}}}$ and in the ZAMO basis $
\boldsymbol{\hat{\mathcal{E}}} = \hat{\mathcal{E}}^{(\alpha)}
\boldsymbol{\hat e}_{(\alpha)}$ reads
\begin{equation}
  \boldsymbol{\hat{\mathcal{E}}}= \frac{\sqrt{\gamma_{\phi\phi}}}{
    \alpha}(\omega_{_{\rm Z}} - \Omega) (\hat{\mathcal{B}}^{(Z)}
  \boldsymbol{\hat e}_{(X)} - \hat{\mathcal{B}}^{(X)} \boldsymbol{\hat
    e}_{(Z)})\,,
\end{equation}
so that it is possible to interpret
$\boldsymbol{\hat{v}}_F = {\sqrt{\gamma_{\phi\phi}}} / \alpha(
\Omega-\omega_{_{\rm Z}}) \boldsymbol{\hat e}_{(Y)}$ as the field lines
velocity in the ZAMO frame, and recover the usual expression
$\boldsymbol{\hat{\mathcal{E}}} = -\boldsymbol{\hat{v}}_F \times
\boldsymbol{\hat{\mathcal{B}}}$.

The orthogonality of the electric and
magnetic fields to each other and to the ZAMO four-velocity implies the
following set of orthonormal vectors~\citep{McKinney:2006sc,
  Chael:2023pwp}
\begin{equation}
\label{eq:TXYZ}
\begin{split}
    {\mathcal{T}}^{(\alpha)}& := {\eta}^{(\alpha)}\,, \quad
    {\mathcal{X}}^{(\alpha)} :=
    \frac{{\hat{\mathcal{E}}}^{(\alpha)}}{\sqrt{\mathcal{E}^2}}\,,\quad
    {\mathcal{Y}}^{(\alpha)} :=
    \frac{{\hat{\mathcal{B}}}^{(\alpha)}}{\sqrt{\mathcal{B}^2}}\,,
    \\ & \quad {\mathcal{Z}}^{(\alpha)} :=
    {\epsilon}^{(\alpha)(\beta)(\gamma)(\delta)}{\mathcal{T}}_{(\beta)}
    {\mathcal{X}}_{(\gamma)}{\mathcal{Y}}_{(\delta)}\,.
\end{split}
\end{equation}
The plasma four-velocity as measured by a ZAMO,
$\boldsymbol{\hat{u}}={\hat{u}}^{(\alpha)}
\boldsymbol{\hat{e}}_{(\alpha)}$ admits the generic decomposition in
terms of components which are parallel, $\parallel$, or orthogonal,
$\perp$, with respect to the magnetic-field lines~\citep{Komissarov2004b,
  McKinney:2006sc, Chael:2023pwp,Gelles2024}
\begin{equation}
  \boldsymbol{\hat{u}} = \hat \gamma (\boldsymbol{\mathcal{T}} +
\hat v_\parallel~\boldsymbol{\mathcal{Y}}+ \hat v_\perp
\boldsymbol{\mathcal{Z}})\,.
\end{equation}

Such a decomposition has the advantage that the spatial velocity
orthogonal to the fields is solely specified by the electromagnetic
sector, $\hat v_\perp := \mathcal{E}/\mathcal{B}$, and the corresponding
Lorentz factor is $\hat \gamma_\perp :=\sqrt{\mathcal{B}^2 /
  (\mathcal{B}^2 - \mathcal{E}^2)}$, where $\mathcal{B}^2 - \mathcal{E}^2
> 0$. The total Lorentz factor $\hat\gamma$ is given by
\begin{equation}
\hat\gamma :=
\frac{1}{\sqrt{1 - \hat{v}_\perp^2 - \hat{v}^2_\parallel}}\,,
\end{equation}
and we note that flows with $\hat \gamma=\hat\gamma_\perp$ (\ie with
$\hat{v}_{\parallel}=0$) provide a unique covariant definition of
timelike observers in FFE characterised by $\Vec{E}\times\Vec{B}$ drift
velocities~\citep{McKinney:2006sc}. Since the total Lorentz factor can
also be expressed as $\hat{\gamma}^2 = \hat{\gamma}_\perp^2/(1 -
\hat{\gamma}^2_\perp \hat{v}_\parallel^2)$, it follows that
$\hat\gamma\geq\hat{\gamma}_\perp$, namely, $\hat\gamma =
\hat{\gamma}_\perp$ for $\hat{v}_\parallel = 0 =: \hat{v}_\parallel^{\rm
  min}$, and $\hat{\gamma}\to\infty$ for $\hat{v}_\parallel =
\hat{\gamma}_\perp^{-1} =: \hat{v}_\parallel^{\rm
  max}$~\citep{Chael:2023pwp}.
%

\subsection{Comoving frame}

Because reconnection is normally studied locally, \ie in the frame
comoving with the plasma~\citep{Liu:2016sqd}, we introduce the comoving
frame with timelike tangent given by the plasma four-velocity
$\boldsymbol{\hat u}$. In the ideal-MHD limit of infinite electrical
conductivity, any electric field is zero in the frame comoving with the
fluid, so that if we define the electric, $\boldsymbol{e}$, and magnetic,
$\boldsymbol{b}$, fields in the comoving frame respectively as $e^\mu :=
F^{\mu \nu} u_\nu$ and $b^\mu := -{^*F}^{\mu \nu} u_\nu$, then $e^\mu =
0$ by construction. Note that while we will employ a comoving frame to
compute the details of the reconnection and plasmoid production, we will
always make use of the electromagnetic fields in the ZAMO frame,
$\boldsymbol{\hat{\mathcal{E}}}$ and $\boldsymbol{\hat{\mathcal{B}}}$, so
that the comoving magnetic field is 
\begin{equation}
  \boldsymbol{b} :=
\frac{\boldsymbol{\hat{\mathcal{B}}} + (\boldsymbol{\hat u}\cdot
  \boldsymbol{\hat{\mathcal{B}}}) \boldsymbol{\hat u}}{\hat \gamma}\,,
\end{equation}
and $b^2:=\boldsymbol{b}\cdot \boldsymbol{b} = \mathcal{B}^2 -
\mathcal{E}^2$.

In the frame comoving with the fluid, we assume that two plasmoids are
ejected locally tangent to the current-sheet with opposite ``outflow''
velocities $\pm \, \Tilde{v}_{\rm out}$, where the signs distinguish the
plasmoid ejected in the direction parallel ($+$) or antiparallel ($-$) to
the {comoving} magnetic-field lines {as seen from one side of the
  reconnection layer}\footnote{{Under realistic conditions as those
  exhibited by the numerical simulations, plasmoids of different sizes
  and and velocities are expected to be produced, so that $\Tilde{v}_{\rm
    out}$ will actually follow a
  distribution~\citep{PhysRevLett.109.265002}.}}. The initial outflow
four-velocity of the plasmoids is thus
\begin{equation}
  \boldsymbol{\tilde{u}}_{\rm
  out}=\tilde \gamma_{\rm out} (\boldsymbol{\hat u}\pm \tilde v_{\rm
  out}~\boldsymbol{b}/b)\,,
\end{equation}
with the azimuthal velocity and Lorentz factor measured by the ZAMO given
by
\begin{align}
  \label{eq:v_out}
  \hat{v}_{\rm out}^{(Y)}& \!=\! \frac{(\hat\gamma_\perp \hat
      v_\parallel \pm \tilde v_{\rm
        out})\hat{\mathcal{B}}^{(Y)}}{\hat\gamma_\perp(1\pm \hat\gamma_\perp
      \hat v_\parallel \tilde v_{\rm out})\mathcal{B}} \!-\! \hat
  v_\perp\frac{\sqrt{\hat{\mathcal{B}}_{(X)}^2+\hat{\mathcal{B}}_{(Z)}^2}}{\mathcal{B}}\,,
  \\
  \label{eq:gamma_out}
    \hat \gamma_{\rm out}& \!=\! \hat{\gamma} \tilde{\gamma}_{\rm out}(1\pm
    \hat{\gamma}_\perp \hat{v}_\parallel \tilde{v}_{\rm out})\,.
\end{align}
%

\subsection{Plasmoids, RAIBs, and PP}

To define the details of the PP we need a local model for magnetic
reconnection that provides a prescription for the velocity field
$\tilde{v}_{\rm out}$ of the plasmoids and their energetics. Such a model
will have to account for the redistribution of the magnetic energy into
the kinetic and internal energy of the plasmoid. The details of this
conversion, which we here assume to originate from a hydrogen plasma, can
be extremely difficult to calculate under realistic conditions, and
probably require a microscopical description involving PIC
simulations~\citep{Ball2018, Meringolo2023}. However, the most important
reconnection quantities that we consider here, namely, the outflow
velocity and the reconnection rate, emerge from MHD constraints and, as
argued by~\citet{Liu:2016sqd}, are not sensitive on the microphysics as
far as reconnection is mediated by plasmoids and the upstream plasma is
in the collisionless regime~\citep{PhysRevD.103.023014}. Approximate
expressions for these quantities are given by
\begin{equation}
  \label{eq: recQ}
  \tilde{v}_{\rm
    out}\approx\sqrt{\sigma_0/(1+\sigma_0)}~\,, \qquad \mathcal{R}_{\rm
    rec}\simeq 0.1~\,,
\end{equation}
with $\sigma_0$ being the upstream plasma magnetisation.

Moreover, to make some analytic progress, we use the so-called
Relativistic Adiabatic Incompressible Ball (RAIB)
model~\citep{Koide:2008xr}, normally adopted under these
conditions~\citep[see][and related works]{Asenjo:2017gsv,
  PhysRevD.103.023014}. Notwithstanding some of their limitations, RAIBs
represent the simplest but non-trivial model of plasmoids, which are
treated as localised distributions of energy dressed with an additional
pressure term that contributes to their inertia and is reminiscent of
their hydrodynamical nature. The RAIB energy-at-infinity needed to
estimate the extraction of energy from a reconnection-driven PP follows
upon integrating the energy-momentum tensor of a perfect fluid and
reads~\citep[see][and also the Appendix]{Koide:2008xr}
\begin{equation}
\label{eq:C3}
    E^\infty=\alpha H \bigg[\hat \gamma_{\rm
        out}\left(1+\frac{\sqrt{\gamma_{\phi\phi}}}{\alpha}\omega_{_{\rm Z}} \hat
      v^{(Y)}_{\rm out}\right)-\frac{U(\Gamma-1)}{H~\hat\gamma_{\rm
          out}}\bigg]\,,
\end{equation}
with $H$, $U$, and $\Gamma$ the total enthalpy, internal energy and
adiabatic index in the plasmoid, respectively. 
\begin{widetext}
  Hence, the energy-at-infinity per unit enthalpy, $\varepsilon^\infty := E^\infty/H$,
is
\begin{align}
\label{eq:E_RAIB_infty}
    \varepsilon^\infty_{\pm}=\alpha \hat{\gamma}&\Bigg[\left(1+\left(
      \hat v_\parallel\frac{\hat{\mathcal{B}}^{(Y)}}{\mathcal{B}}-\hat v_\perp
     \frac{\sqrt{\hat{\mathcal{B}}_{(X)}^2+\hat{\mathcal{B}}_{(Z)}^2}}{\mathcal{B}}\right)\frac{\sqrt{\gamma_{\phi\phi}}}{\alpha}
      \omega_{_{\rm Z}}\right)\tilde{\gamma}_{\rm out}+ \\ \nonumber
      &~~\pm\left(\hat v_\parallel \hat
      \gamma_\perp+\left(\frac{\hat{\mathcal{B}}^{(Y)}}{\hat \gamma_\perp\mathcal{B}}-
      \hat v_\parallel \hat v_\perp\hat
      \gamma_\perp\frac{\sqrt{\hat{\mathcal{B}}_{(X)}^2+\hat{\mathcal{B}}_{(Z)}^2}}{\mathcal{B}}\right)\frac{\sqrt{\gamma_{\phi\phi}}}
            {\alpha}\omega_{_{\rm Z}}\right)\tilde{v}_{\rm
              out}\tilde{\gamma}_{\rm
              out}-\frac{U(\Gamma-1)/H}{\hat\gamma^2 \tilde{\gamma}_{\rm
                out}\left(1\pm \hat\gamma_\perp \hat v_\parallel
              \tilde{v}_{\rm out}\right)}\Bigg]\,.
\end{align}
\end{widetext}

\begin{figure}[h!]
  \centering
  \includegraphics[width=0.95\columnwidth]{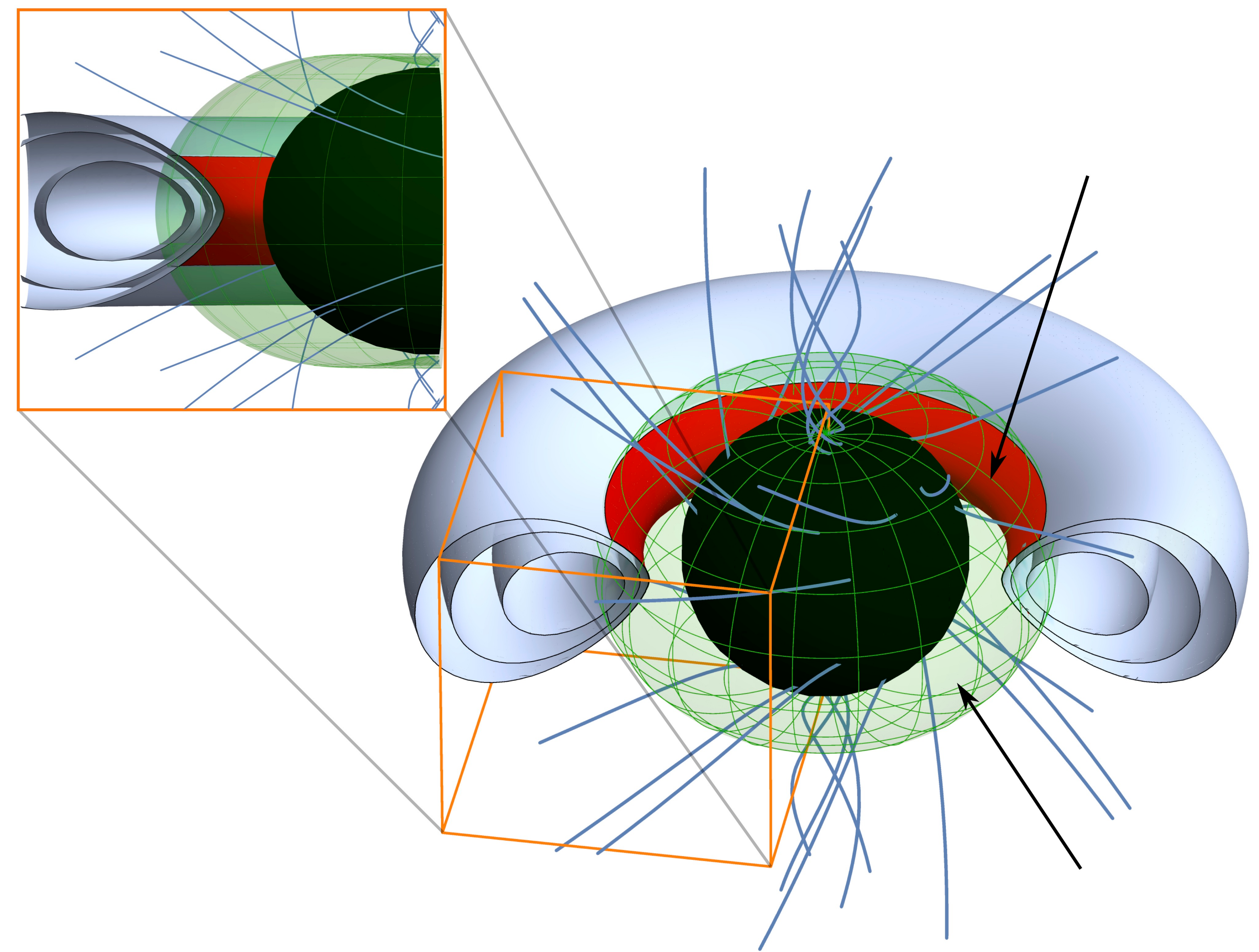}
    \begin{picture}(0,0)
        \put(-40,148){``${\rm ergobelt}$''}
        \put(-40,8){${\rm ergosphere}$}
    \end{picture}
 \caption{Shown with grey contours are the iso-levels of the rest-mass
   density of a torus with a toroidal magnetic field entering the BH
   ergosphere (green region). Reconnection and plasmoid production can
   take place on the surface of the torus and if this happens in the
   ergoregion, \ie on the ``ergobelt'' (red region), then a PP can be
   activated from plasmoids having a negative energy-at-infinity.}
  \label{fig:nicePic}
\end{figure}
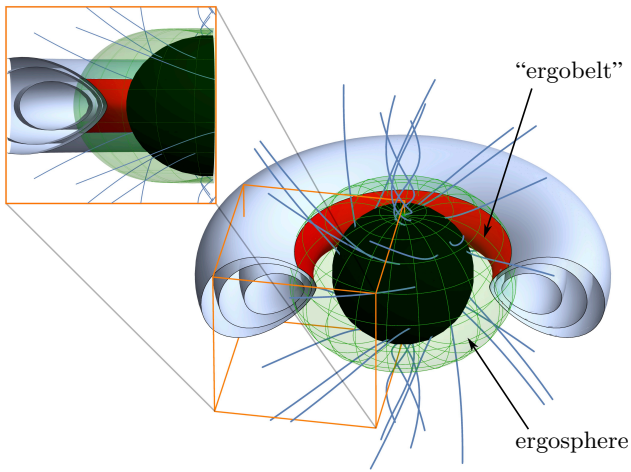

Expression~\eqref{eq:E_RAIB_infty} contains information not only on the
background geometry via the metric functions $\alpha$, $\omega_{_{\rm
    Z}}$, and $\gamma_{\phi\phi}$, but also on the global magnetic-field
configuration around the BH, either via the magnetic field components
$\hat{\mathcal{B}}^{(\alpha)}$, or through the field strength given by
$\hat v_\perp = \mathcal{E} / \mathcal{B}$. In addition, the dynamics of
the bulk plasma is encoded in $\hat v_\parallel$, that should be
prescribed together with the coordinate invariant quantities
$\Psi(r,\theta)$, $I(r,\theta)$, and $\Omega(\Psi)$. Note that
Eq.~\eqref{eq:E_RAIB_infty} is consistent with any stationary and
axisymmetric solution in ideal MHD and in FFE, and can be used to model
BH-energy extraction by plasmoids in more realistic configurations. By
contrast, previous models were restricted to the equatorial plane and
invoked a Keplerian dynamics not realistic within the
ergosphere~\citep{Koide:2008xr, Asenjo:2017gsv, PhysRevD.103.023014}.

\section{Application to a Magnetised torus}

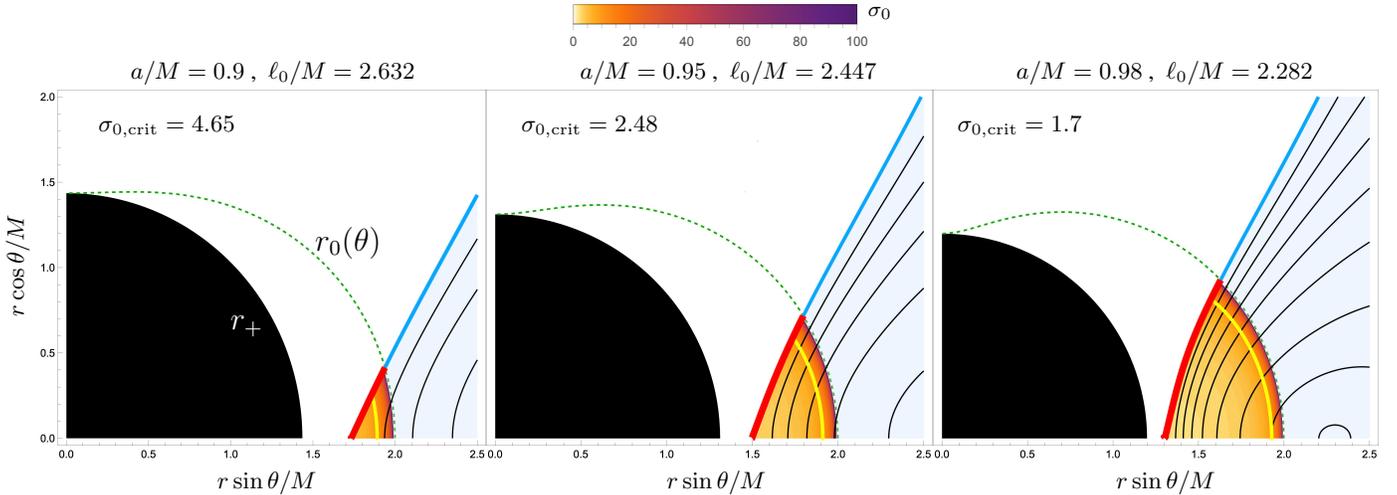
\begin{figure*}
    \centering \includegraphics[width=0.25\textwidth]{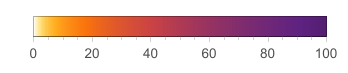} \\
    \begin{picture}(0,0)
      \put(60,27){\footnotesize$\sigma_0$}
    \end{picture}
    \\
    \vspace{0mm}
    \includegraphics[width=0.33\textwidth]{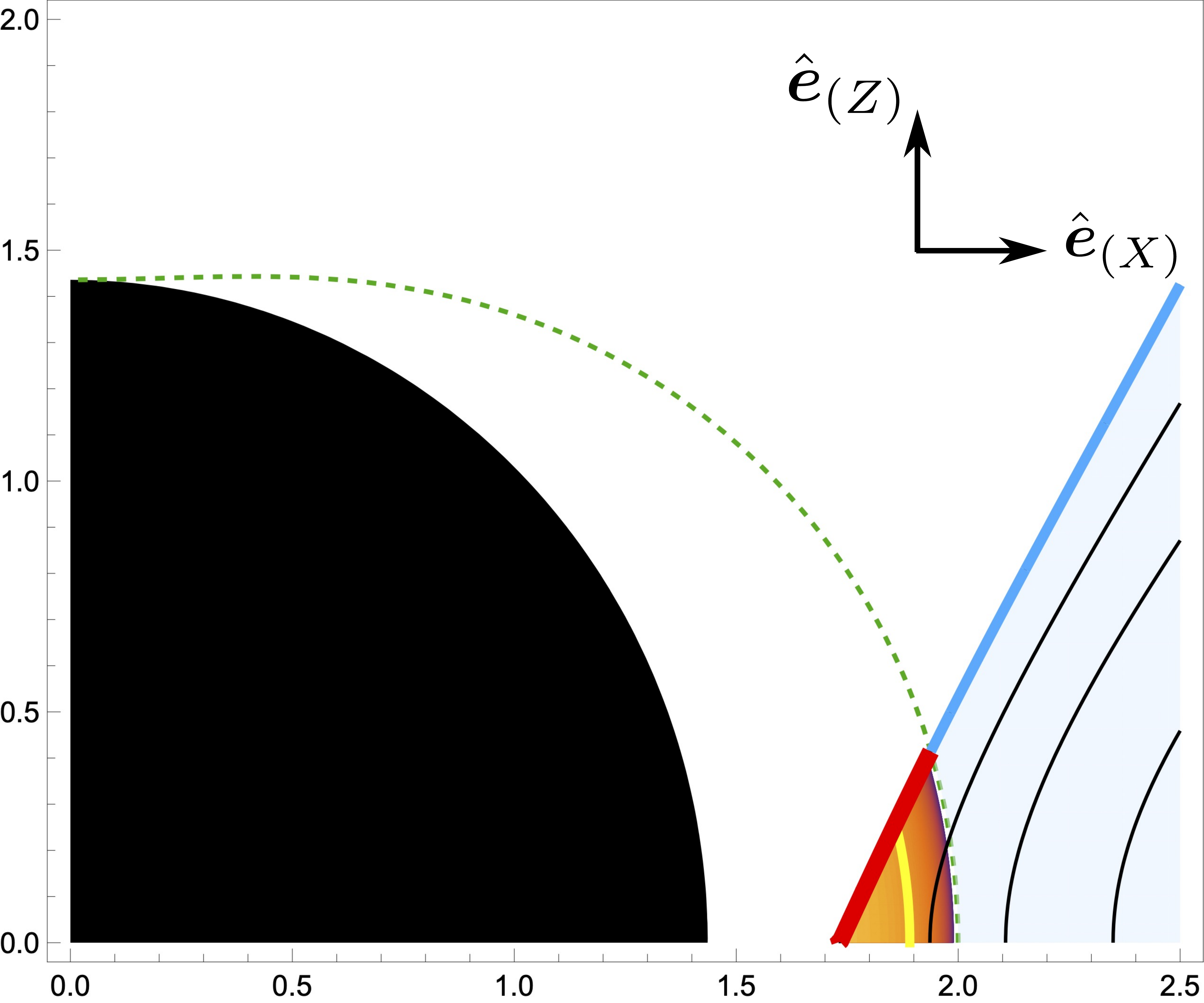}
    \hspace{-0.6mm}\includegraphics[width=0.33\textwidth]{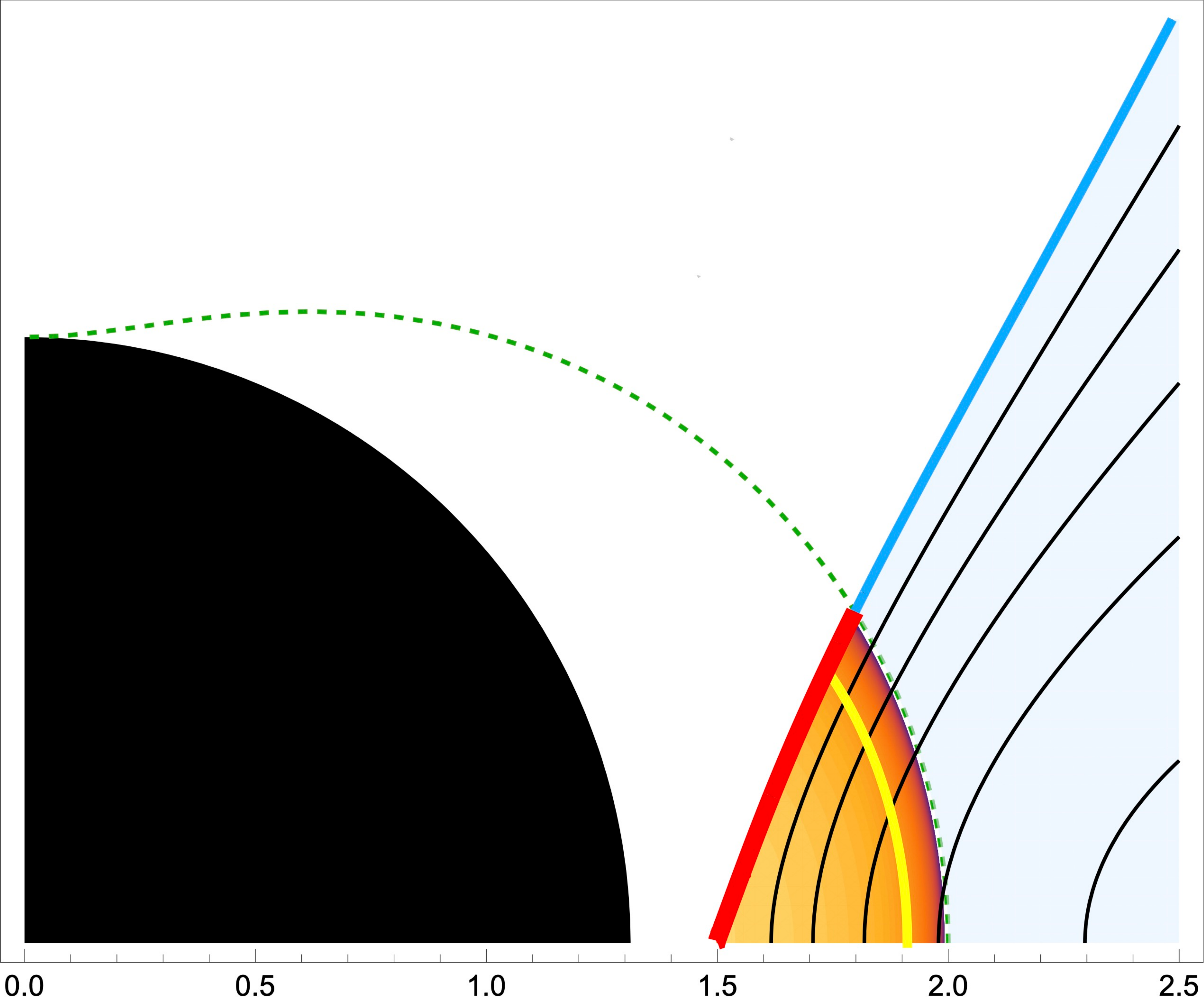}
    \hspace{-0.6mm}\includegraphics[width=0.33\textwidth]{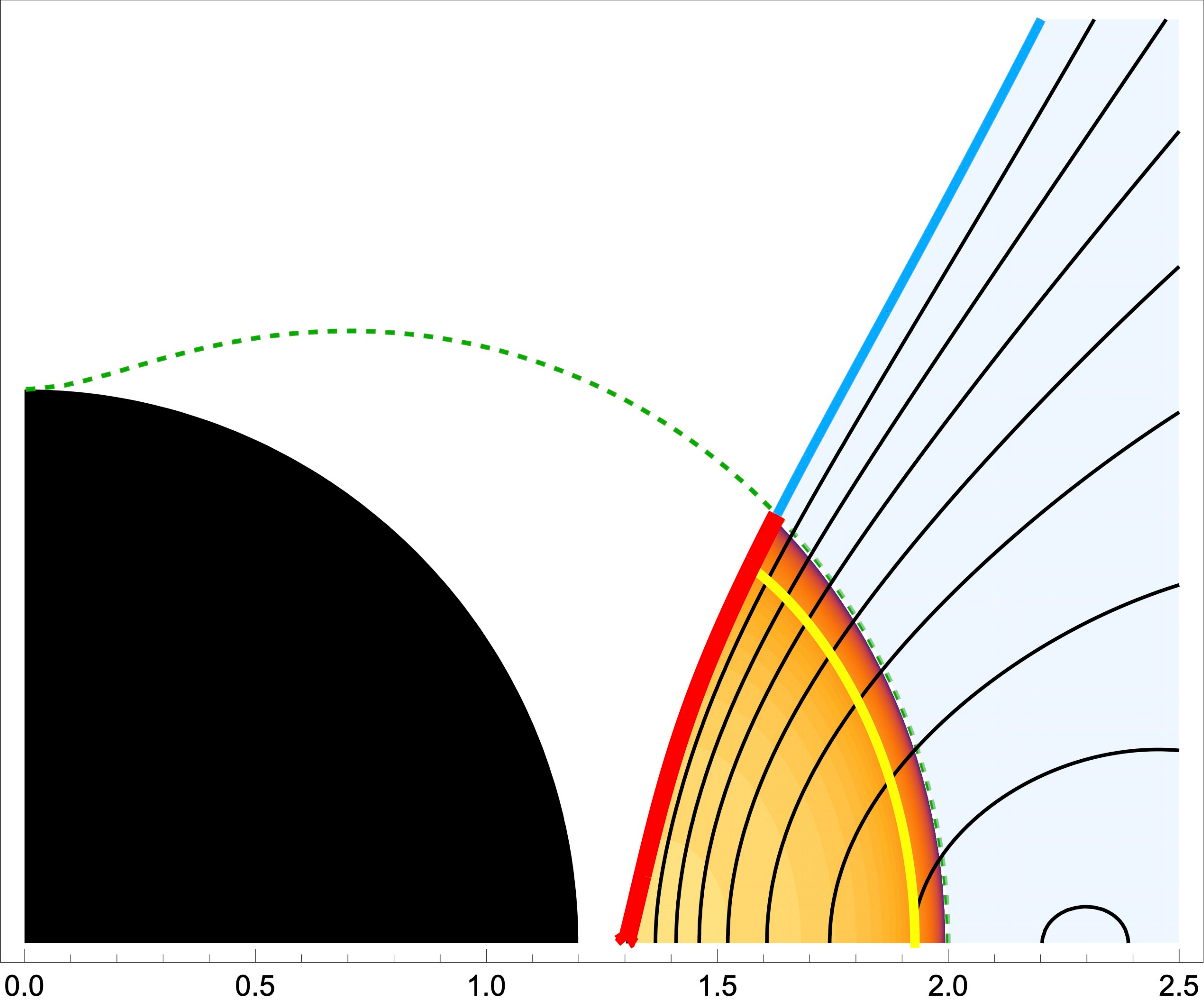}
    \\
    \begin{picture}(0,0)
      \put(-185,0){\small$r\sin\theta/M$}
      \put(-15,0){\small$r\sin\theta/M$}
      \put(155,0){\small$r\sin\theta/M$}
      \put(-220,155){\small $~a/M=0.9~$, $~\ell_0/M=2.632$}
      \put(-50,155){\small $~a/M=0.95~$, $~\ell_0/M=2.447$}
      \put(115,155){\small $~a/M=0.98~$, $~\ell_0/M=2.282$}
      \put(-230,135){\small $\sigma_{0, {\rm crit}}=4.65$}
      \put(-70,135){\small $\sigma_{0, {\rm crit}}=2.48$}
      \put(95,135){\small $\sigma_{0, {\rm crit}}=1.7$}
      \put(-265,65){\rotatebox{90}{\small$r\cos\theta/M$}}
       \put(-180,60){\textcolor{white}{\large$r_+$}}
      \put(-148,90){\large$r_0(\theta)$}
    \end{picture}
    \vspace{1mm}
    \caption{Polar sections of tori with $\ell_0 \simeq \ell_{\rm mb}$
      (light blue shading) entering the ergosphere $r_0$ (green dashed
      line), with BH spin increasing from left to right. The torus
      surface inside the ergosphere, dubbed ergobelt (red solid line), is
      suitable for reconnection to drive a PP. The colormap reports
      values of the magnetisation $\sigma_0$ necessary for the
      pair-production of plasmoids with $\varepsilon^\infty_+>0$ and
      $\varepsilon^\infty_-<0$ at different positions in the torus, with
      the yellow solid line marking $\sigma_0=10$.}
    \label{fig:Disc_models}
\end{figure*}

All studies to date of reconnection-driven PP have been restricted to
scenarios where current-sheets develop on the BH equatorial
plane. However, there is growing numerical evidence, either from
MHD~\citep{Mahlmann2020, Ripperda2020, Nathanail2020, Nathanail2020c,
  Ripperda2022, Dimitropoulos:2024iqz} or from PIC
simulations~\citep{ElMellah:2021tjo, ElMellah:2023sun,Vos2024}, that
significant reconnection takes place also in the transition zone between
the thick accretion disc and the ensuing jet launched by the BH. More
specifically, such simulations have shown that the accretion disc around
BHs develops a toroidal component and that the transition region between
the torus and the funnel systematically leads to reconnection
phenomena.

To achieve a treatment in which the off-equatorial dynamics of
matter is not prescribed ``by hand'' (as done so far with equatorial
current sheets) but follows self-consistently from a stable fluid
configuration, we apply our general treatment to well-known Komissarov's
solution, namely, to a geometrically thick disc endowed with purely
toroidal magnetic fields corotating with the BH~\citep{Komissarov2006a,
  Montero07, Gimeno-Soler:2017}.
The presence of an inner light surface will naturally lead to dominant
toroidal magnetospheric fields in the ergoregion with polarity in the
counterrotating direction~\citep{Komissarov2004b, Uzdensky2005,
  Nathanail2014, Gralla:2014yja, Camilloni:2022kmx}. Indeed, a necessary
condition for the BZ mechanism to occur is $\omega_{_{\rm{BH}}} >
\Omega>0$~\citep{Lasota:2013kia}. This implies that the field lines
tension oppose to the BH rotation, so that the azimuthal components of
the magnetic field must have polarity in the counter-rotating direction,
$\hat{\mathcal{B}}^{(Y)} = \sqrt{\gamma_{\phi\phi}}
\hat{\mathcal{B}}^\phi\propto -I \propto -
(\omega_{_{\rm{BH}}}-\Omega)<0$ close to the horizon~\citep[see, for
  instance,][for a derivation]{Camilloni:2022kmx}.

Hence, with the simultaneous presence of toroidal magnetic fields with
corotating polarity within the torus and toroidal magnetospheric fields
with counterrotating polarity within the funnel, the accretion-disc
surface will develop a current-sheet and act as an optimal site for the
production plasmoids, as also found in numerical
simulations~\citep{Dimitropoulos:2024iqz}\footnote{Under realistic
conditions of accretion and outside the ergosphere, the magnetic field
will also have a poloidal component both in the torus and in the jet, so
that the presence of guide fields can be relevant for reconnection taking
place far from the ergosphere.}. Furthermore, the portion of the torus
surface that is inside the ergosphere and undergoes reconnection, \ie
what we dub \emph{``ergobelt''} (red region in Fig.~\ref{fig:nicePic}),
can generate plasmoids that have negative energy-at-infinity and hence
tap the BH rotational energy via a plasmoid-driven PP.
Note that the current sheet on the torus surface will
also be accompanied by the standard equatorial current-sheet that inevitably
develops for split-field configurations; we refer to~\citet{Koide:2008xr,
  Asenjo:2017gsv, PhysRevD.103.023014, Parfrey2019, Chen:2024ggq} to
review the phenomenology and energetics associated to these current
sheets.

Such a magnetised torus, thus, provides a simplified but non-trivial
configuration that is mathematically self-consistent and does not invoke
unrealistic assumptions. The Komissarov solution naturally entails that
$\Psi=0$ and
\begin{equation}
  \hat{v}_\parallel =\ell_0 \frac{\alpha}{ \sqrt{\gamma_{\phi\phi}}
  (\omega_{_{\rm Z}}\ell_0-1)}\,,
\end{equation}
so that $\hat v_\perp=0$ and
$\hat{\mathcal{B}}^{(X)}=\hat{\mathcal{B}}^{(Z)}=0$ (see
Appendix). Hereafter we will consider maximally-filled tori with
$\ell_0\approx \ell_{\rm mb}$ and $\mathcal{W}_{\rm in}=\mathcal{W}_{\rm
  cusp}=0$.

Following~\citet{Koide:2008xr} and~\citet{PhysRevD.103.023014}, we assume
the plasmoids/RAIBs to consist of relativistically-hot plasma described
by an ideal-fluid equation of state so that $H=4U(\Gamma-1)$.  As
mentioned above, and since we are interested in the PP associated with an
accretion scenario onto a supermassive black hole, the reconnection we
envisage to take place at the ergobelt is one that involves a hydrogen
plasma. Moreover, we use the analytic model of magnetic reconnection
by~\citet{Liu:2016sqd},
with a reconnection electric field proportional to the reconnection rate,
and predicting plasmoid velocities that are simple functions of the
magnetisation $\sigma_0 := b^2/w_0$, where $w_0$ is the enthalpy
density. More specifically, we set~\citep{Lyubarsky2005}
\begin{equation}
\tilde{v}_{\rm
  out}\approx\sqrt{\sigma_0/(1+\sigma_0)}\,, \qquad \tilde{\gamma}_{\rm
  out}\approx\sqrt{1+\sigma_0}\,, 
\end{equation}
so that Eq.~\eqref{eq:E_RAIB_infty} simplifies and becomes dependent on
the BH spin $a$, the plasma magnetisation\footnote{Typical magnetisation
just inside the torus is $\sim\mathcal{O}(0.01-0.1)$ with vanishing
magnetic fields at the surface. Numerical simulations show that
$\sigma_0$ jumps to $\sim\mathcal{O}(10-100)$ just outside the
torus~\citep{Mahlmann2020, Ripperda2020, Nathanail2020, Nathanail2020c,
  Ripperda2022, Dimitropoulos:2024iqz}. Hence, to model the reconnection
in the transition region in the vicinity of the ergobelt we consider $0.1
\lesssim \sigma_0 \lesssim 10$.} $\sigma_0$, the torus angular momentum
$\ell_0$, and the $(r,\theta)$ location of the ergobelt (see
Eq.~\eqref{eq:eRAIB_VII} in the Appendix) .

A number of important remarks should be made on
$\varepsilon_{\pm}^\infty$ as obtained above. First, the value at the
horizon, $\varepsilon_{\pm}^\infty\big\vert_{r_+}$, is finite for all BH
spins. Second, since the conditions for a successful PP require
$\varepsilon^\infty_+ > 0$ and $\varepsilon^\infty_- <
0$~\citep{PhysRevD.103.023014}, a critical magnetisation $\sigma_{0, {\rm
    crit}}$ exists that depends on the BH spin and torus properties and
below which no PP can take place. For instance, for $a/M=0.9$ and
$\ell_0/M \lesssim 2.63$, $\sigma_{0, {\rm crit}} \approx 4.65$, which is
consistent with the values observed in the transition region in
simulations~\citep{Nathanail2020, Nathanail2020c}. Third, in regions
where $\sigma_0 \gg 1$, Eq.~\eqref{eq:E_RAIB_infty} further simplifies to
\begin{equation}
  \varepsilon^\infty_\pm = \sqrt{\sigma_0}\left(1\pm
  \frac{\sqrt{\gamma_{\phi\phi}}}{\alpha} \omega_{_{\rm
      Z}}\right)\sqrt{\frac{1\pm \hat v_\parallel}{ 1 \mp \hat
      v_\parallel }}\,,
\end{equation}

showing that plasmoids with $\varepsilon^{\infty}_{-}<0$ can be produced
for sufficiently high magnetisation, provided that a portion of the torus
intersects the ergosphere where $\omega_{_{\rm Z}}
\sqrt{\gamma_{\phi\phi}} \geq \alpha$. For maximally filled tori, the
torus cusp is inside the ergosphere if
\begin{equation}
a > a_{\rm crit} := 2(\sqrt{2}-1)\, M \approx 0.83 \, M\,.
\end{equation}
Finally, and most importantly, because Eq.~\eqref{eq:E_RAIB_infty}
applied to a torus is not restricted to the equatorial plane, it allows
us to study for the first time the problem plasmoid-driven PP in an
axisymmetric but non-equatorial context.

Figure~\ref{fig:Disc_models} shows polar sections of magnetised tori with
$\ell_0 \lesssim \ell_{\rm mb}$ entering the ergosphere of BHs with
spin. Marked with a colormap are the values of the magnetisation
$\sigma_0$ such that $\varepsilon^\infty_+>0$ and
$\varepsilon^\infty_-<0$ [\cf Eq.~\eqref{eq:E_RAIB_infty}] and reported
are corresponding the critical value for the magnetisation increasing
from left to right. For magnetisations above a critical value $\sigma_{0,
  {\rm crit}}$, the innermost part of the torus possesses an ergobelt as
a site for the production of plasmoids with negative energy-at-infinity
and hence yield a PP. Moreover, for faster spinning BHs, larger portions
of the torus penetrate the ergosphere increasing the potential production
of plasmoids.

\subsection{Energetics of the plasmoid-driven PP}

Because of the nonlinear nature of reconnection and of the plasma
dynamics near the torus surface, the energy extraction process will be
intrinsically stochastic, so that the energy-extraction rate
$\dot{E}_{\rm rec}$ will depend on the local reconnection rate, routinely
assumed to be $\mathcal{R}_{\rm rec} \simeq 0.1$~\citep{Liu:2016sqd,
  Cassak:2017enb, PhysRevLett.109.265002, Asenjo:2017gsv,
  PhysRevD.103.023014}. Taking now $\mathcal{L}$ and $\mathcal{A}$
respectively as the linear size and surface area of the ergobelt (note
there are two ergobelts, symmetric with respect to the equatorial plane),
we approximate the enthalpy density as $w_0\approx H/ (\mathcal{L}
\mathcal{A})$ and express the energy extraction rate as
\begin{equation}
  \dot{E}_{\rm
  rec} = -2\,\mathcal{R}_{\rm rec}\, w_0 \, \int_{\mathcal{A}}
\varepsilon_-^\infty d\mathcal{A} = -2\, \mathcal{R}_{\rm rec}\, w_0 \,
\mathcal{A} \, \langle \varepsilon_-^\infty \rangle\,,
\end{equation}
where
\begin{equation}
\langle \varepsilon_-^\infty \rangle := \frac{\int_{\mathcal{A}}
\varepsilon_-^\infty d\mathcal{A}}{\int_{\mathcal{A}} d\mathcal{A}}\,,
\end{equation}
is the energy-at-infinity
averaged over the area of the ergobelt (see Appendix).

In the top part of Fig.~\ref{fig:EnExtr} we report the behaviour of
$\dot{E}_{\rm rec}/w_0$ as a function of the plasma magnetisation and BH
spin. As expected, no energy extraction occurs if the BH is not spinning
sufficiently fast (the torus does not enter the ergosphere) or if the
magnetisation is below the corresponding critical value (red solid
line). Conversely, the process is more efficient for rapidly spinning BHs
and high magnetisations, since the effective area of the ergobelt then
increases. The values of $\dot{E}_{\rm rec}$ are comparable or larger
than those estimated by~\citet{PhysRevD.103.023014} for plasmoids
generated in the equatorial plane.

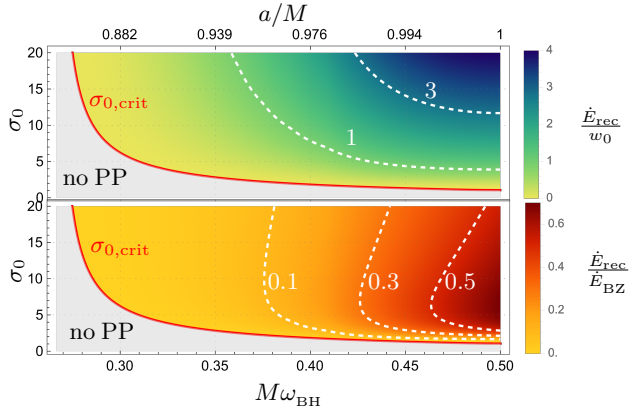
\begin{figure}
  \vspace{2mm}\hspace{-10mm} \centering
  \includegraphics[width=0.9\columnwidth]{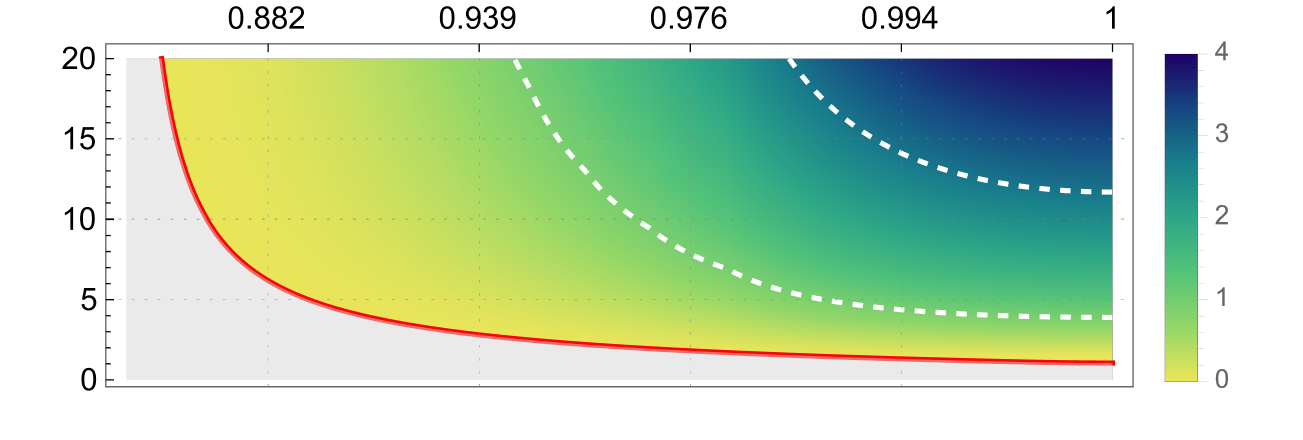}
     \begin{picture}(0,0)
       \put(-126,77){$a/M$}
       \put(-5,35){{{$\frac{\dot{E}_{\rm rec}}{w_0}$}}}
       \put(-200,15){${\rm no~PP}$}
       \put(-188,45){\textcolor{red}{$\sigma_{0, {\rm crit}}$}}
       \put(-220,35){\rotatebox{90}{$\sigma_0$}}
       \put(-93,30){\textcolor{white}{\small{$1$}}}
       \put(-63,48){\textcolor{white}{\small{$3$}}}
    \end{picture}
    \\
    \vspace{-4.7mm}\hspace{-7.2mm}
    \includegraphics[width=0.899\columnwidth]{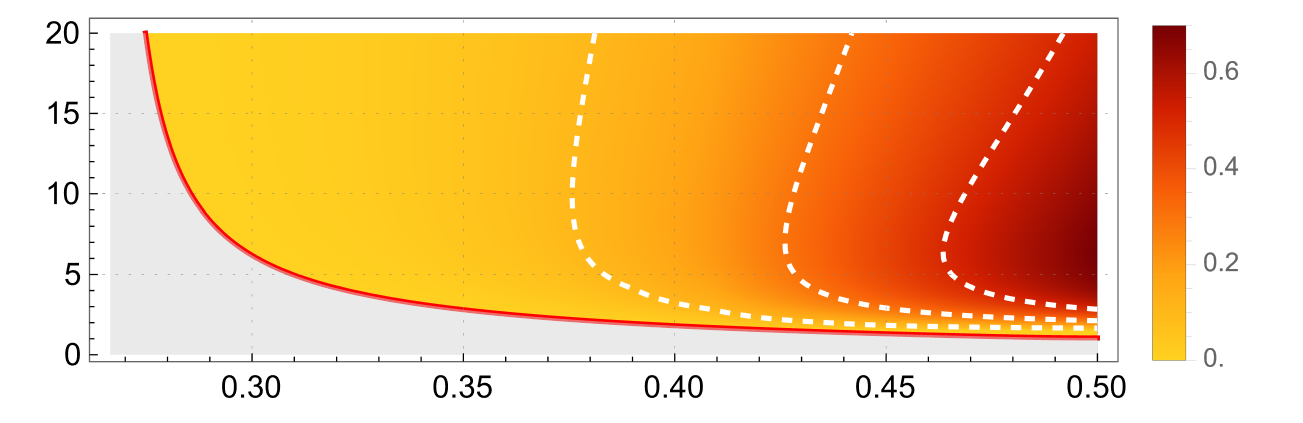}
    \begin{picture}(0,0)
      \put(-5,40){{{$\frac{\dot{E}_{\rm rec}}{\dot{E}_{_{\rm BZ}}}$}}}
      \put(-221,35){\rotatebox{90}{ $\sigma_0$}}
      \put(-200,16){${\rm no~PP}$}
      \put(-192,49){\textcolor{red}{$\sigma_{0, {\rm crit}}$}}
      \put(-130,-7){$M\omega_{_{\rm BH}}$}
      \put(-123,35){\textcolor{white}{\small{$0.1$}}}
      \put(-86,35){\textcolor{white}{\small{$0.3$}}}
      \put(-57,35){\textcolor{white}{\small{$0.5$}}}
    \end{picture}
    \caption{\textit{Top panel:} Rate of energy extraction per unit of
      enthalpy density, ${\dot{E}_{\rm rec}}/w_0$, at the ergobelt of a
      maximally-filling torus as a
      function of the magnetisation $\sigma_0$ and BH angular velocity
      $\omega_{_{\rm BH}}$ (the top horizontal axis uses instead a scale
      in $a/M$). The red solid line marks the critical magnetisation
      $\sigma_{0, {\rm crit}}$ below which no PP is possible (grey shaded
      area). \textit{Bottom panel:} The same as in the top but when the
      energy extraction rate is normalised to the BZ power ${\dot{E}_{\rm
          rec}}/{\dot{E}_{_{\rm BZ}}}$.}
    \label{fig:EnExtr}
\end{figure}

Finally, we can compare the extracted energy via plasmoids with the
power extracted via the BZ mechanism, which, we recall, is steady and
effective for any nonzero value of the BH spin. The BZ power extracted
by a split-monopole field in Kerr can be written as an expansion in
the BH angular velocity, $\omega_{_{\rm BH}} := \omega_{_{\rm
    Z}}\vert_{r_+}$, to ensure a better
convergence~\citep{Tchekhovskoy:2009ba, Camilloni:2022kmx}, and reads
\begin{equation}
  \label{eq:EBZ}
  \dot{E}_{_{\rm BZ}} \!\simeq\! \left(\frac{2\pi B_0^2 r_+^2
  \omega_{_{\rm BH}}^2}{3} \right) \!\left[1 \!+\!
  1.38\,(M\omega_{_{\rm BH}})^2 \!-\!  9.2\,(M\omega_{_{\rm
      BH}})^4\right]\,,
\end{equation}
where, for simplicity, we assumed that the magnetic field at the event
horizon is comparable to that at the ergobelt, \ie $B^2_{r=r_+}
\approx B^2_0 = \sigma_0 w_0$, and where the magnetic flux
  across the horizon is given by $\Phi_{\rm H}\approx 4\pi B_0r_+^2$.
The bottom part of Fig.~\ref{fig:EnExtr} reports the ratio
$\dot{E}_{\rm rec}/\dot{E}_{_{\rm BZ}}$ showing that for sufficiently
large BH spins and magnetisations, the plasmoid-driven PP leads to an
energy extraction that is smaller but comparable to that associated
with the BZ mechanism.
  Figure~\ref{fig:EnExtr} invites a comparison with Fig. (7) in
  \citet{PhysRevD.103.023014}, where $\dot{E}_{\rm rec}/\dot{E}_{\rm
    BZ}$ was also computed assuming plasmoids generated from an
  equatorial current sheet and having Keplerian velocities. When
  making such a comparison, we should take into account differences in
  the physical regimes considered and in the approximations made. In
  particular, since we consider an accretion scenario, we limit the
  magnetisation to regimes that are compatible with the simulations
  and hence consider $\sigma_0 \lesssim 20$
  [\citet{PhysRevD.103.023014} consider instead $\sigma_0 \lesssim
    10^6$]. Furthermore, the magnetic flux considered by
  \citet{PhysRevD.103.023014} includes a factor proportional to
  $\sin\xi$, where $\xi$ is the ``orientation angle'', that is, the
  angle between the radial and azimuthal directions of the plasmoid
  velocity as measured in the comoving frame. The inclusion of this
  angular dependence, which is unsual for a surface-integrated flux
  that should not depend on the orientation angle, decreases the BZ
  power by a factor $\sin^2\xi$, thus magnifying the ratio by $\sim
  15$ for $\xi=\pi/12$ [which is the value considered in Fig.~(7)
    of~\citet{PhysRevD.103.023014}]. When ignoring this factor and
  considering similar values of magnetisation, our estimates of
  $\dot{E}_{\rm rec}/\dot{E}_{\rm BZ}$ are comparable to those
  computed by \citet{PhysRevD.103.023014} and by \citet{Parfrey2019}
  for electrons undergoing a PP.

\section{Conclusions}
\label{sec:conclusions}

Responding to the increasing evidence that resistive effects in plasmas
around BHs play an important role in shaping the observed phenomenology
of these objects, we have presented a general approach to the study of a
PP driven by plasmoids produced at reconnection sites along current
sheets. Our formalism ultimately establishes the physical conditions to
make a plasmoid-driven PP energetically viable and can be applied equally
to scenarios that are dominated by the plasma or by the magnetic field,
that is, in MHD or FFE regimes. It is also genuinely multidimensional and
hence allows one to explore conditions that are beyond the ones studied
so far and that have been restricted to the equatorial plane. In this
sense, it provides a direct contact with numerical simulations, either in
MHD~\citep{Mahlmann2020, Ripperda2020, Nathanail2020, Nathanail2020c,
  Ripperda2022, Dimitropoulos:2024iqz} or with PIC
approaches~\citep{Parfrey2019, ElMellah:2021tjo,
  ElMellah:2023sun,Vos2024}, all of which highlight an intense
reconnection activity outside the equatorial plane. Finally, it does not
assume an ad-hoc description of the dynamics of the plasma, or conjecture
its kinematic properties with oversimplified and possibly unrealistic
configurations. On the contrary, it constructs the dynamics of the plasma
starting from a well-known configuration, that of an equilibrium torus
with a toroidal magnetic field, and thus possessing all the features
necessary to compute self-consistently the reconnection process and
estimate the PP energetics.

While the results presented here offer the first coherent approach to a
multidimensional treatment of the PP driven by magnetic reconnection,
they can be improved in a number of ways. First, by developing a more
accurate descriptions of plasmoids that could overcome the limitations of
the RAIB model. Second, by exploiting the results of numerical
simulations to produce better estimates of the reconnection rate and
hence of the efficiency of the energy-extraction process. Thirdly, by
connecting the plasmoid production at the ergobelt with the plasmoid
production on the equatorial plane (that remains a favourable site for
the production of plasmoids), hence join the MHD regime of the plasma in
the torus with the FFE regime when the plasma has left the torus and is
accreting onto the BH. We plan to investigate these important aspects in
future works.

\section*{Acknowledgements}

We thank Beno\^{i}t Cerutti, Luca Comisso, Ileyk El Mellah, Claudio
Meringolo, Antonios Nathanail, and Kyle Parfrey for insightful
discussions and comments. Partial funding comes from the State of Hesse
within the Research Cluster ELEMENTS (Project ID 500/10.006), by the ERC
Advanced Grant ``JETSET: Launching, propagation and emission of
relativistic jets from binary mergers and across mass scales'' (Grant
No. 884631). LR acknowledges the Walter Greiner Gesellschaft zur
F\"orderung der physikalischen Grundlagenforschung e.V. through the Carl
W. Fueck Laureatus Chair.


\newpage  
\appendix 

In what follows we provide the most important details about the
calculations or considerations that are needed to obtain the results
presented in the main text.

\section{Kerr metric and ZAMO frame.~}
\label{Sec:3+1}
\begin{figure}[h!]
    \centering
    \includegraphics[width=0.25\textwidth]{Legend.pdf}
    \\
    \begin{picture}(0,0)
        \put(60,25){\footnotesize$\sigma_0$}
    \end{picture}
    \\
     \vspace{-5mm}
    \includegraphics[width=0.5\columnwidth]{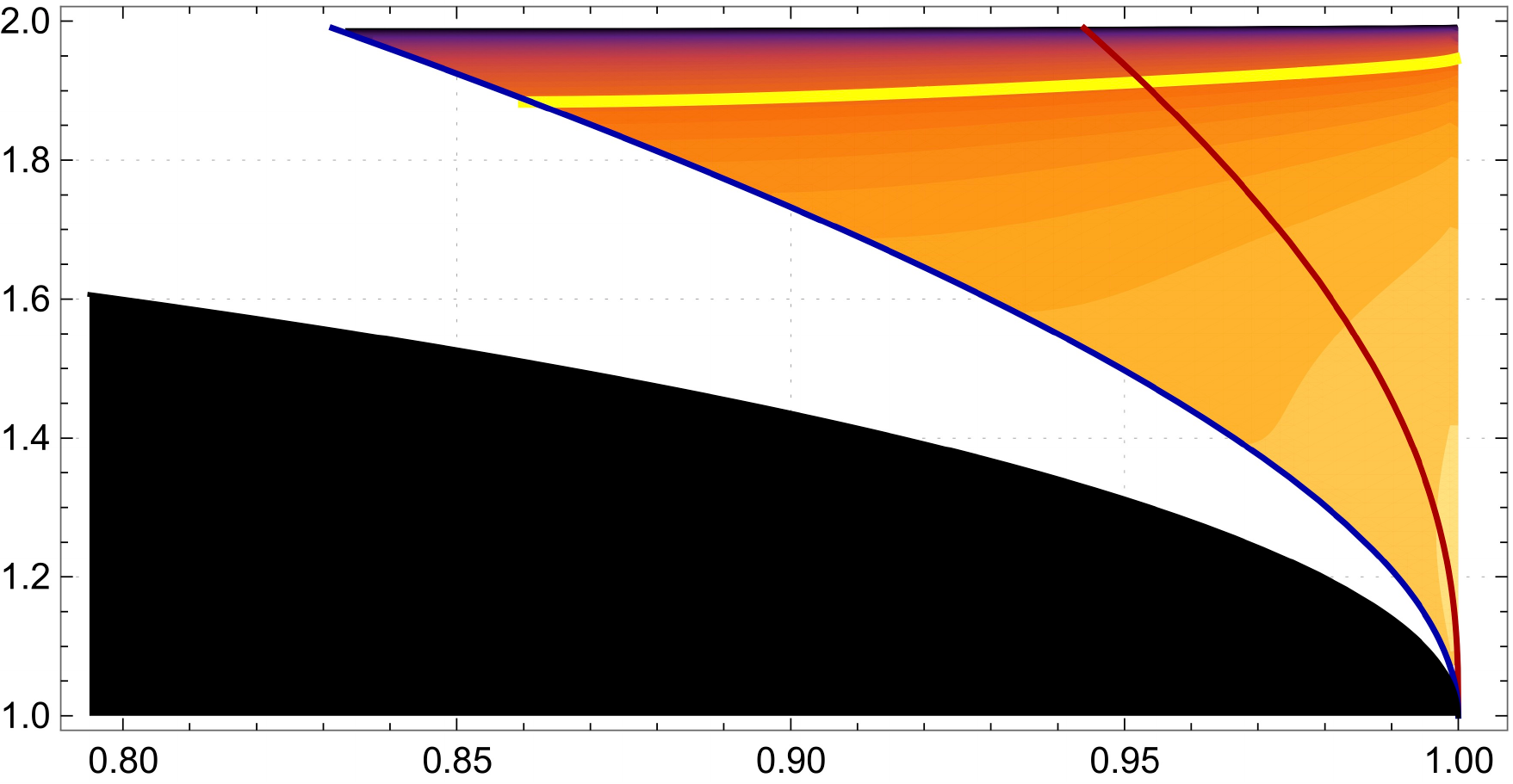}
     \begin{picture}(0,0)
        \put(-229,49){\rotatebox{90}{$r/M$}}
        \put(-120,-10){ $a/M$}
        \put(-120,45){\textcolor{white}{\large$r_+$}}
        \put(-130,80){\large$r_{\rm mb}$}
        \put(-33,80){\large$r_{\rm ms}$}
    \end{picture}

\vspace{1mm} \caption{The same as Fig.~\ref{fig:Disc_models} but for
    $\theta=\pi/2$ and illustrating the $(r,a)$-dependence. For each BH
    spin, the torus cusp is located between $r_{\rm mb}$ (black line) and
    $r_{\rm ms}$ (red line). Maximally filled tori penetrate the
    ergosphere ($r_0=2\,M$ on the equator) for $a/M \gtrsim 0.83$. Tori
    with the cusp at the largest possible position ($r_{\rm cusp} =
    r_{\rm ms}$ or $\ell_0 = \ell_{\rm ms}$) require larger BH spins,
    $a/M \gtrsim 0.94$. By increasing the BH spin $r_{\rm ms} \to r_{\rm
    mb} \to r_{+}$, the ergobelt area grows and comparatively smaller
    magnetisations are needed for a plasmoid-mediated
    PP.}  \label{fig:PLR}
\end{figure}

The Kerr metric in Boyer-Lindquist (BL) coordinates
$x^\mu=(t,r,\theta,\phi)$ and in a $(3+1)$-form reads
$ds^2=g_{\mu\nu}dx^\mu dx^\nu=-\alpha^2 dt^2+\gamma_{ij}(dx^i+\beta^i
dt)(dx^j+\beta^j dt)$ with the {lapse function}, the {shift vector} and
the {absolute space metric} respectively defined as
\begin{align}
    &\alpha=\frac{1}{\sqrt{-g^{tt}}}=\sqrt{\frac{\Delta \Sigma}{\Pi}}\,,
    \quad \beta^i=\alpha^2
    g^{ti}=-\alpha^2 \frac{2M a r}{\Delta \Sigma}~\delta^i_\phi\,, \quad\gamma_{ij} dx^i
    dx^j={\Sigma}/{\Delta}\,dr^2
    + \Sigma\,d\theta^2+{\Pi}/{\Sigma}\sin^2\theta\,d\phi^2\,,
\end{align}
and the metric determinant decomposes as $g
=- \alpha^2\det(\gamma_{ij})$. The metric functions explicitly read
$\Delta = r^2-2M r +a^2$, $\Sigma = r^2+a^2\cos^2\theta$ and $\Pi =
(r^2+a^2)^2-a^2\Delta \sin^2\theta$, with the event horizon position
given by the $r_+$ root of $\Delta=(r-r_+)(r-r_-)$, $r_\pm =
M(1 \pm \sqrt{1-a^2_*})$, and the ergosphere location at $r_0(\theta) =
M(1 + \sqrt{1-a^2_*\cos^2\theta})$, where $a_* := a/M$. We assume the BH
to have $0 \leq a_* \leq 1$, with the upper bound imposed by the cosmic
censorship conjecture. Other interesting locations in the BH equatorial
plane, $\theta=\pi/2$, are given by the {marginally stable} corotating
($+$) and counter-rotating ($-$) circular orbit, $r_{\rm ms,\pm}$, and by
the {marginally bound} circular orbit $r_{\rm ms,\pm}$, depicted in Fig.~\ref{fig:PLR}.

The ZAMO is a normal observer (the corresponding $1$-form in BL
coordinates is $\boldsymbol{\eta} = -\alpha~ \boldsymbol{dt}$), with an
associated local Cartesian right-handed frame
$\boldsymbol{\hat{e}}_{(\alpha)} = (\boldsymbol{\hat{e}}_{(T)},
\boldsymbol{\hat{e}}_{(X)}, \boldsymbol{\hat{e}}_{(Y)},
\boldsymbol{\hat{e}}_{(Z)})$ specified by the vectors
$\boldsymbol{\hat{e}}_{(T)} = \boldsymbol\eta$, $\boldsymbol{\hat
  e}_{(X)}={\boldsymbol{\partial_r}} / {\sqrt{\gamma_{rr}}}$,
$\boldsymbol{\hat e}_{(Y)} = \boldsymbol{\partial_\phi} /
\sqrt{\gamma_{\phi\phi}}$, and $\boldsymbol{\hat
  e}_{(Z)}=-{\boldsymbol{\partial_\theta}} /
\sqrt{\gamma_{\theta\theta}}$, and form a tetrad, namely, $g_{\mu\nu} =
\hat{e}^{(\alpha)}_\mu \hat{e}^{(\beta)}_\nu
\hat{\eta}_{(\alpha)(\beta)}$ and $\varepsilon_{\mu\nu\rho\sigma} = \hat
e^{(T)}_\mu\wedge \hat e^{(X)}_\nu \wedge \hat e^{(Y)}_\rho \wedge \hat
e^{(Z)}_\sigma$, where $\hat{\eta}_{(\alpha)(\beta)}$ is the Minkowski
metric in the local frame and $\varepsilon_{\mu\nu\rho\sigma} :=
\sqrt{-g}\eta_{\mu\nu\rho\sigma}$ is the $4$-dimensional Levi-Civita
tensor.

\section{Plasma velocity.~}
As mentioned in the main text, the character of the field is set by the
invariant quantity $ F^{\mu\nu}F_{\mu\nu}=2
(\mathcal{B}^2-\mathcal{E}^2)$, with the strength of the ZAMO fields
given by
\begin{equation}
\label{eq:B2E2}
\begin{split}
  \mathcal{B}^2&=\frac{1}{\sin^2\theta}\left[\frac{\Delta (\partial_r
      \Psi)^2+(\partial_\theta\Psi)^2}{\Pi}+\frac{I^2}{\Delta}\right]\,,\quad
  \mathcal{E}^2=\mathcal{F}^2(\Omega)\frac{\Delta (\partial_r
      \Psi)^2+(\partial_\theta\Psi)^2}{\Pi\sin^2\theta}\,,
\end{split}
\end{equation}
and where we introduced $\mathcal{F}(\Omega) := \sqrt{\gamma_{\phi\phi}}
  (\omega_{_{\rm Z}} - \Omega) / \alpha$. Notice that
  $\mathcal{F}(\Omega) = \pm 1$ at the outer/inner light surfaces, \ie
  the locations where corotating observers with the field lines become
  null, $|\partial_t + \Omega(\Psi)
\partial_\phi|^2 = \alpha^2(\mathcal{F}^2(\Omega)-1) = 0$, or
equivalently, where the magnetic field transitions from being dominated
by poloidal to toroidal components.

The orthonormal vectors (see Eq.~\eqref{eq:TXYZ} in the main text), valid
for degenerate fields, are conveniently parameterised in terms of two
magnetic orientation angles $\xi_1 := \tan^{-1}( \hat{\mathcal{B}}^{(Y)}
/ \hat{\mathcal{B}}^{(X)} )$ and $\xi_2 := \cos^{-1}
({\hat{\mathcal{B}}}^{(Z)} / \sqrt{\mathcal{B}^2})$ with respect to the
ZAMO frame, such that
\begin{align}
\label{eq:sinxi}
    \sin\xi_1&=-{\sqrt{\Pi}I}/{\sqrt{\Delta
        (\partial_\theta\Psi)^2\! + \!\Pi I^2}}\,,
        \\
        \sin\xi_2&=\sqrt{({\Pi I^2\! + \!\Delta(\partial_\theta\Psi)^2})/({\Pi
        I^2\! + \!\Delta\left( \Delta
        (\partial_r\Psi)^2\! + \!(\partial_\theta\Psi)^2\right))}}\,, \nonumber
 \end{align}
with explicit expressions given by
\begin{equation}
\begin{split}
	&\boldsymbol{\mathcal{X}}=\mathcal{X}^{(\alpha)}\boldsymbol{\hat
    e}_{(\alpha)}=({\cos\xi_2}/\chi)\,\boldsymbol{\hat
    e}_{(X)}-({\cos\xi_1\sin\xi_2}/\chi)\,\boldsymbol{\hat e}_{(Z)} \\
    &\boldsymbol{\mathcal{Y}}=\mathcal{Y}^{(\alpha)}\boldsymbol{\hat
    e}_{(\alpha)}=\cos\xi_1\sin\xi_2 \,\boldsymbol{\hat
    e}_{(X)}+\sin\xi_1\sin\xi_2 \,\boldsymbol{\hat
    e}_{(Y)}+\cos\xi_2\,\boldsymbol{\hat e}_{(Z)}\,, \\
    &\boldsymbol{\mathcal{Z}}={\mathcal{Z}}^{(\alpha)}\boldsymbol{\hat
    e}_{(\alpha)}=-({\cos\xi_1\sin\xi_1\sin^2\xi_2}/\chi)\,\boldsymbol{\hat
    e}_{(X)} + \chi\,\boldsymbol{\hat
    e}_{(Y)}-({\cos\xi_2\sin\xi_1\sin\xi_2}/\chi)\,\boldsymbol{\hat e}_{(Z)}\,,
\end{split}
\end{equation}
where $\chi:= \sqrt{1-(\sin\xi_1\sin\xi_2)^2}$.

As anticipated in the main text, it is possible to use such a set of
vectors to construct the most general four-velocity that is compatible
with the electric-field screening condition. More specifically, the
velocity perpendicular to the fields can be written as $\hat{v}_\perp^2 =
\mathcal{F}^2(\Omega) (1 - \hat{\mathcal{B}}_{(Y)}^2/ \mathcal{B}^2)$.
The parallel component, instead, can be parametrized as
$\hat{v}_\parallel:=q \hat{v}_\parallel^{\rm max}=q/\hat{\gamma}$ with
$q\in(-1,1)$ and $\hat\gamma=\hat\gamma_{\perp}/\sqrt{1-q^2}$.
In the ZAMO Cartesian frame, the components of the spatial velocity read
\begin{align}
  \label{eq:vxvyvz_1}
    \hat{v}^{(X)}& =
    \left(\mathcal{F} \frac{\hat{\mathcal{B}}^{(Y)}}{\mathcal{B}}
    \!+\! q \sqrt{1 \!-\! \mathcal{F}^2 \left(1 \!-\!
      \frac{\hat{\mathcal{B}}_{(Y)}^2}{\mathcal{B}^2}\right)}\right)
    \frac{\hat{\mathcal{B}}^{(X)}}{\mathcal{B}}\,, \\
\label{eq:vxvyvz_2}
    \hat{v}^{(Y)}& =
    \mathcal{F} \left(\frac{\hat{\mathcal{B}}_{(Y)}^2}{\mathcal{B}^2}
    \!-\! 1\right) \!+\! q \sqrt{1 \!-\! \mathcal{F}^2 \left(1 \!-\!
      \frac{\hat{\mathcal{B}}_{(Y)}^2}{\mathcal{B}^2}\right)}
    \frac{\hat{\mathcal{B}}^{(Y)}}{\mathcal{B}}\,, \\
\label{eq:vxvyvz_3}
    \hat{v}^{(Z)}& =
    \left(\mathcal{F} \frac{\hat{\mathcal{B}}^{(Y)}}{\mathcal{B}}
    \!+\! q \sqrt{1 \!-\! \mathcal{F}^2 \left(1 \!-\!
      \frac{\hat{\mathcal{B}}_{(Y)}^2}{\mathcal{B}^2}\right)}\right)
    \frac{\hat{\mathcal{B}}^{(Z)}}{\mathcal{B}}\,,
\end{align}
where $0\leq q \leq 1$ is a free coefficients and where the expressions
above reduce to the previous results in the literature under the
appropriate limits\footnote{For $q=0$,
expressions~\eqref{eq:vxvyvz_1}--\eqref{eq:vxvyvz_3} coincide with (D140)
in~\citet{Chael:2023pwp}, and for $q=1$ with (A12) in
~\citet{Komissarov2004b}. Furthermore, the following relations hold:
$\hat{\mathcal{B}}^{(X)} = \sqrt{\gamma_{rr}}\hat{\mathcal{B}}^r$,
$\hat{\mathcal{B}}^{(Y)}
= \sqrt{\gamma_{\phi\phi}}\hat{\mathcal{B}}^\phi$ and
$\hat{\mathcal{B}}^{(Z)} =
- \sqrt{\gamma_{\theta\theta}} \hat{\mathcal{B}}^\theta\,.$}
Furthermore, it is easy to verify that the components~\eqref{eq:vxvyvz_1}
and \eqref{eq:vxvyvz_3} satisfy the standard relation
$ \hat{v}^{(X)}/ \hat{\mathcal{B}}^{(X)}=
\hat{v}^{(Z)}/\hat{\mathcal{B}}^{(Z)}$.
%

\section{On the magnetised torus.~}
Equations~\eqref{eq:vxvyvz_1}--\eqref{eq:vxvyvz_3} show that if the
magnetic field is purely toroidal, $\boldsymbol{\hat{\mathcal{B}}}
= \hat{\mathcal{B}}^{(Y)} \boldsymbol{\hat{e}}_{(Y)}$, the same is true
for the plasma velocity $\boldsymbol{\hat v}
=q \, \boldsymbol{\hat{e}}_{(Y)}$, so that the velocity is purely
parallel to the field lines. This is the basic dynamical property of the
plasma in a Komissarov torus.
Furthermore, the condition for the four-velocity to be that of a circular
motion with angular velocity $\Omega_0$ as seen by an asymptotic observer
requires that $\hat{v}_\parallel = q = \mathcal{F}(\Omega_0)$.
We here are interested in tori with a constant specific angular-momentum
$\ell_0 := u_{\phi}/u_t$ and corotating with the BH ($\ell_0 = {\rm
const.}  >0$), so that the magnetic field is aligned with the azimuthal
coordinate, \ie $\Psi=0$ and $I<0$, so that $\xi_{1,2}=\pi/2$ [see
Eq.~\eqref{eq:sinxi}]. In this case, $\Omega_0 :=
-(g_{t\phi}+g_{tt}\ell_0) / (g_{\phi\phi}+g_{t\phi}\ell_0)$, and we can
express the parallel velocity as a function of $\ell_0$ as
$\hat{v}_\parallel={\alpha}{\ell_0}/[{\sqrt{\gamma_{\phi\phi}}}
({\omega_{_{\rm Z}}\ell_0-1})]$.
With these assumptions, together with the condition $\tilde{v}_{\rm
out}\approx\sqrt{\sigma_0/(1+\sigma_0)}$ and $\tilde{\gamma}_{\rm
out}\approx\sqrt{1+\sigma_0}$~\citep{Liu:2016sqd}, Eq.~\eqref{eq:C3}
reduces to
\begin{align}
  \label{eq:eRAIB_VII}
\varepsilon^\infty_{\pm} = &\alpha \hat{\gamma}\left[\left(1 +
\frac{\sqrt{\gamma_{\phi\phi}}}{\alpha}\omega_{_{\rm
  Z}} ~\hat{v}_\parallel\right)\sqrt{1+\sigma_0}
    \pm\left(\hat{v}_\parallel+\frac{\sqrt{\gamma_{\phi\phi}}}{\alpha}
    \omega_{_{\rm Z}}\right)\sqrt{\sigma_0}-\frac{1}{4}\frac{\sqrt{1+\sigma_0}
      - \hat{v}_\parallel \sqrt{\sigma_0}}{\hat{\gamma}^2
      \left(1+\sigma_0-\hat{v}_\parallel^2\sigma_0\right)}\right]\,.
\end{align}
We should note that although we derive Eq.~\eqref{eq:eRAIB_VII} as a
limit of Eq.~\eqref{eq:C3} for $\hat{v}_\parallel =
{\alpha}{\ell_0}/[{\sqrt{\gamma_{\phi\phi}}} ({\omega_{_{\rm Z}} \ell_0 -
1})]$ and $\tilde{v}_{\rm out}\approx\sqrt{\sigma_0 / ( 1 +\sigma_0)}$,
it coincides with Eq.~(34) in~\citet{PhysRevD.103.023014}, where the
plasma was assumed to be in a Keplerian circular orbit in the equatorial
plane. Within these limits, therefore, Eq.~(34)
in~\citet{PhysRevD.103.023014} would provide a multidimensional expression
for the plasmoid energies when the Keplerian velocity is replaced by a
generic nonplanar one.

We notice that the comoving magnetic field satisfies $b^t=\ell_0b^\phi$,
as typical in the Komissarov torus~\citep{Komissarov2004b}, with
$b^t=-\hat{v}_\parallel\hat{\gamma}_{\parallel}
\hat{\mathcal{B}}^{(Y)}/\alpha$\footnote{In the notation used in
~\citet{Komissarov2004b}, $b^\phi = \pm\sqrt{2p_m/\mathcal{A}}$, where
$p_m := b^2/2 = \hat{\mathcal{B}}^2_{(y)}/2$ and $\mathcal{A}:=
g_{\phi\phi} + 2\ell_0 g_{t\phi} + \ell_0^2g_{tt} =
\alpha^2 \ell_0^2 (1 - \hat{v}_\parallel^2) / \hat{v}_\parallel^2$.}.
The torus is traditionally defined by equipotential surfaces specified by
the condition~\citep{Abramowicz78, Kozlowski1978}
\begin{equation}
  \label{eq:W_potential}
  \begin{split}
      \mathcal{W}&=\frac{1}{2}\ln\left\vert\frac{\alpha^2\gamma_{\phi\phi}}
       {\gamma_{\phi\phi}(1-\ell_0\omega_{_{\rm Z}})^2-\alpha^2\ell_0^2}
       \right\vert={\rm const.}\,.
  \end{split}
\end{equation}
We label with $\mathcal{W}_{\rm cusp}$ the equipotential surface
intersecting the equatorial plane at the corresponding {cusp} radius
$r_{\rm cusp}$, for which the angular momentum matches the Keplerian
value $\ell_0=\ell_{_{\rm K}}$, and whose value is solely specified in
terms of $\ell_0$ and $a$, whereas $\mathcal{W}_{\rm in}$ labels the
torus inner edge, that must be specified as a boundary
condition. Furthermore, a choice for $\ell_0$ uniquely determines the
locations of the cusp and of the position of the (pressure) rest-mass
density maximum and is constrained to be $\ell_0>\ell_{\rm
ms}$~\citep{Rezzolla_book:2013}.
If $\ell_{\rm ms} \leq \ell_0 \leq \ell_{\rm mb}$, then $\mathcal{W}_{\rm
cusp}$ is guaranteed to be non-positive, so that matter and fields can
fill the torus up to the cusp, $\mathcal{W}_{\rm in}\leq\mathcal{W}_{\rm
cusp}$. In the limiting case $\ell_0=\ell_{\rm mb}$, the outermost
equipotential surface is marginally closed, \ie $\mathcal{W}_{\rm
cusp}=0$, while if $\mathcal{W}_{\rm in}=0$ the torus is said to be
maximally filled. Conversely, if $\ell_0>\ell_{\rm mb}$, then
$\mathcal{W}_{\rm cusp}>0$, and matter never reaches the cusp being
confined within the regions where $\mathcal{W}_{\rm in}<0$.  In our
construction, we always consider tori that are maximally filled,
$\mathcal{W}_{\rm in}=\mathcal{W}_{\rm cusp}=0$, since this arguably
represents the most interesting configuration, with the torus
configuration fully specified by the BH spin and with matter at the cusp
prone to accrete onto the BH if perturbed.

The area of the ergobelt is determined by an implicit equation
$\mathcal{W}_{\rm in}(r,\theta) = \mathcal{W}_0$ . Because of the
non-trivial location of the ergobelt, the surface differential
$d\mathcal{A}$ is given by $ d\mathcal{A} = \sqrt{\gamma_{\phi\phi}
d\phi^2} \sqrt{\gamma_{\theta\theta} d\theta^2 + \gamma_{rr}dr^2}
\Big\vert_{\mathcal{W}_0}$, with $\phi\in[0,2\pi]$ and
$\theta\in[\theta_\star, \pi/2]$, where $\theta_\star$ is the polar angle
that limits the effective region from above at given $\sigma_0$, $\ell_0$
and $a$. The minimum value $\theta_{\star}^{\rm m}$ (\ie the maximum of
the integral support, corresponding to the optimal condition
$\sigma_0\gg1$) can be determined analytically by inverting
Eq.~\eqref{eq:W_potential} with the supplementary condition
$r=r_0(\theta_\star^{\rm m})$, so that 
\begin{align}
    &\cos(2\theta^{\rm m}_\star)\!=\!1\! + \!\frac{4 f_0 \ell_0}{a_* M
    (1\!  - \!2f_0)}\! + \!\frac{4f_0}{{a^2_*(1\! - \!2f_0)^2}}\Big(f_0\!
  - \!1\! + \!  \sqrt{(f_0\! - \!1)^2\! - \!{a^2_*}(1\!  - \!2f_0)^2\! +
    \!2{a_* \ell_0} f_0(2f_0\!  - \!1)/{M}}~\Big)\,,
\end{align}
where $f_0 := e^{2\mathcal{W}_0}$. For a maximally filled torus, $f_0=1$,
and $\cos(2\theta^{\rm m}_\star) = 1 - 4 \ell_0
/a+4M\sqrt{2\ell_0-a}/a^{3/2}$, which consistently leads to $\theta^{\rm
m}_\star=\pi/2$ for $a_{\rm crit}=2(\sqrt{2}-1)M$.

\section{Relativistic Adiabatic Incompressible Balls (RAIBs).~}
In what follows we recall the calculation of the energy-at-infinity of a
RAIB starting from the energy momentum tensor of an ideal fluid,
following the original discussion of Koide and Arai~\citep{Koide:2008xr}.
The hydrodynamic energy and angular momentum as measured by a ZAMO can be
obtained by projecting the fluid part of the energy-momentum tensor
$T^{\mu\nu}_{\rm fl}$ (\ie not containing electromagnetic contributions)
over the ZAMO tetrad
\begin{equation}
    \hat e_{\rm fl}\! := \!T_{\rm fl}^{(T)(T)}\! = \!-p+w \hat \gamma^2\,,
    \quad
    \hat \ell_{\rm fl}\! := \!T_{\rm fl}^{(T)(Y)}\! = \!w \hat \gamma^2\hat v^{(Y)}\,,
\end{equation}
where $T^{\mu\nu}_{\rm fl} := w u^\mu u^\nu+pg^{\mu\nu}$, with $w$ and
$p$ the enthalpy density and the pressure measured in the comoving frame
of the fluid, and $\hat\gamma$ the Lorentz factor of the fluid with
respect to the ZAMO. The conserved Noether currents in stationary and
axisymmetric hydrodynamic flows, $\mathcal{J}^\mu_E =
-T^{\mu}_{~~\nu}(\partial_t)^\nu$ and $\mathcal{J}^\mu_L =
T^{\mu}_{~~\nu}(\partial_\phi)^\nu$, can be expressed as combinations of
quantities measured in the ZAMO frame, with their redshifted temporal
components being respectively the energy density and
angular-momentum-at-infinity, respectively $e^{\infty}=\alpha \hat{e} +
\sqrt{\gamma_{\phi\phi}} \omega_{_{\rm Z}} \hat \ell$, and $\ell^\infty =
\sqrt{\gamma_{\phi\phi}}~\hat \ell$.

Assuming the fluid to be perfect and described by an ideal-fluid equation
of state, the pressure can be expressed as $p = \rho \epsilon (\Gamma -
1)$, with $\Gamma$ the adiabatic index, $\rho$ the rest-mass density, and
$\epsilon$ the specific internal energy. As a result, the enthalpy
density is $w=\rho(1 + \epsilon \Gamma) = \rho + 4 p$ for a completely
degenerate relativistic fluid with
$\Gamma={4}/{3}$~\citep{Rezzolla_book:2013}. For simplicity, we assume
that the rest-mass density of the plasmoid is much smaller than the
internal energy density, \ie $\rho \ll \rho \epsilon$, so that the
enthalpy density can be approximated as $w= \rho + 4 p \approx
4p$. Future improvement to the approach followed here (and in the
literature) will concentrate on better estimating not only the role of
the rest-mass density over the internal energy density, but also on the
role played by the magnetisation in the plasma.

Bearing these considerations in mind, we recall that the RAIB model
assumes that the rest-mass density is localised, so that, in the ZAMO
frame, $\rho = {m} \delta^3 (\boldsymbol{\hat{x}} -
\boldsymbol{\hat{x}}(t)) / {\hat{\gamma}(t)}$, where $m$ is the rest-mass
of the RAIB. This allows to easily integrate the energy and angular
momentum density over the entire three-space in the ZAMO frame to obtain
the total energy and angular-momentum-at-infinity associated to a RAIB
\begin{align}
\label{eq:ERAIB}
E^\infty_{_{\rm RAIB}}
  &=\alpha\left[H\hat \gamma\left(1
  + \frac{\sqrt{\gamma_{\phi\phi}}}{\alpha}\omega_{_{\rm Z}} \hat
  v^{(Y)}\right)- \frac{U(\Gamma-1)}{\hat \gamma}\right]\,,\quad
  L^\infty_{_{\rm RAIB}} =\sqrt{\gamma_{\phi\phi}}~H \hat{\gamma} \hat{v}^{(Y)}\,.
\end{align}
In obtaining the expressions above we have written the total enthalpy as
$H=m(1 + \epsilon \Gamma)$ and the total internal energy as $U=m\epsilon$
[Eq.~\eqref{eq:ERAIB} coincides with Eq.~\eqref{eq:C3} in the main text].

We conclude this section by clarifying an inconsistency often encountered
in the literature that adopts the RAIB prescription. It should be noted
that in~\citet{PhysRevD.103.023014} [see Eq.~(25) therein], and in
many related works, the expression above is improperly identified as an
energy density. The inconsistency becomes evident once realised that if
Eq.~(25) in \citet{PhysRevD.103.023014} corresponds to an energy density,
it can be further integrated in space to obtain the total energy, and
upon taking the point-like particle limit (\ie $U=0$ and $H=m$) in flat
spacetime ($\alpha=1$ and $\omega_{_{\rm Z}}=0$), this would lead to
$E=m$. Conversely, the point-like particle limit in flat spacetime of
Eq.~\eqref{eq:C3} leads to the correct result for the total relativistic
energy of a single particle, \ie $E=m\hat{\gamma}$.

\section*{Data Availability}
All data are incorporated into the article and its online Supporting
Information.

\end{document}